\documentclass{article}

\usepackage{amssymb,amsfonts,amsmath}
\usepackage{cite,enumerate,float}
\usepackage{color}
\usepackage{tikz}
\usetikzlibrary{arrows,snakes,backgrounds}

\usepackage[utf8]{inputenc}

\def\be{\begin{eqnarray}}
\def\ee{\end{eqnarray}}

\definecolor{red}{rgb}{1,0,0}
\definecolor{orange}{rgb}{1,0.5,0}
\definecolor{violet}{rgb}{0.7,0,1}

\hoffset= - 1.0in         

\begin{document}

\hfill MIPT/TH-06/26

\hfill IITP/TH-06/26

\hfill ITEP/TH-06/26

\bigskip

\centerline{\Large{
Correlators in the theory of Integral Discriminants
}}

\bigskip

\centerline{A. Morozov$^{a,b,c,d}$, Sh. Shakirov$^c$}

\bigskip

\begin{center}
$^a$ {\small {\it MIPT, Dolgoprudny, 141701, Russia}}\\
$^b$ {\small {\it NRC ``Kurchatov Institute", 123182, Moscow, Russia}}\\
$^c$ {\small {\it Institute for Information Transmission Problems, Moscow 127994, Russia}}\\
$^d$ {\small {\it Institute for Theoretical and Experimental Physics, 117218, Moscow, Russia}}
\end{center}

\bigskip

\centerline{ABSTRACT}

\bigskip

{\footnotesize

Integral discriminants provide a simple and fundamental model for non-Gaussian integrals, associated with homogeneous polynomials of degree $r$ in $n$ variables. We argue that, in this context, the study of correlators is equally if not more important. In this paper, we study a natural class of correlators in this model -- the invariant correlators. We suggest a general method to compute invariant correlators using differential operators that act on the partition function. This method allows to compute general invariant correlators in terms of the fundamental invariants. Moreover, in some cases the correlators appear to be simply polynomials in the invariants. This could be an interesting manifestation of superintegrability phenomenon in the theory of integral discriminants.

}

\bigskip

\bigskip

\section{Introduction}

Non-Gaussian integrals present a significant challenge in theoretical physics. Since Gaussian integrals are very well understood, non-Gaussian integrals are often treated approximately as close to Gaussian. This gives rise to a powerful set of methods known as perturbation theory. However, perturbation theory has its limitations and many interesting phenomena -- the non-perturbative effects -- stay beyond its reach. Therefore, it is important to develop exact (non-perturbative) methods that can deal with non-Gaussian integrals.

An interesting approach is provided by the study of \emph{integral discriminants} \cite{IntDisc} (an earlier paper closely related to this subject is \cite{QFT3}) -- simplest non-Gaussian integrals that can be constructed from homogeneous polynomials $S$ of degree $r$ in $n$ variables:

\begin{align}
Z_{n|r}(S) = \int\limits_{\gamma} d^n x e^{-S(x_1, \ldots, x_n)}, \ \ \ \ \ S(x_1, \ldots, x_n) = \sum\limits_{i_1, \ldots, i_r = 1}^{n} S_{i_1, \ldots, i_r} x_{i_1} \ldots x_{i_r},
\end{align}
\smallskip\\
where the contour of integration $\gamma$ can be chosen arbitrarily. For $r = 2$ this is just a Gaussian integral, 

\begin{align}
Z_{n|2}(S) = \int\limits_{\gamma} d^n x e^{-\sum S_{ij} x_i x_j} \simeq \dfrac{1}{\sqrt{\det S}},
\end{align}
\smallskip\\
however for $r > 2$ this model is non-Gaussian and much more interesting. In a sense, for $r > 2$ it is fully non-perturbative because the Gaussian term in the exponent is fully absent, only the non-Gaussian term is present. An important feature is that, just like for $r = 2$, the singularities of $Z_{n|r}(S)$ for $r > 2$ are situated along the locus of algebraic discriminant \cite{GKZ, NewAndOld} $D_{n|r}(S) = 0$ (of which $\det S = 0$ is the simplest example). This was the reason why we suggested the term \emph{integral discriminants} for the functions $Z_{n|r}(S)$. Altogether, this can be considered as a part of the larger \emph{non-linear algebra} program \cite{NonLinearAlgebra1, NonLinearAlgebra2}.

\pagebreak

In \cite{IntDisc}, we listed the simplest properties of integral discriminants, including

\paragraph{$\bullet$ $SL(n)$ invariance.}Integral discriminant is an invariant function of $S$:

\begin{align}
Z_{n|r}({\widetilde S}) = Z_{n|r}(S), \ \ \ {\widetilde S}_{i_1, \ldots, i_r} = \sum\limits_{j_1, \ldots, j_r = 1}^{n} S_{j_1, \ldots, j_r} U_{i_1, j_1} \ldots U_{i_r, j_r},
\end{align}
\smallskip\\
where $U \in SL(n)$. As a corollary, it can be expressed through fundamental polynomial invariants of $S$:

\begin{align}
Z_{n|r}(S) = Z_{n|r}\big(\mbox{ Inv}_1(S), \mbox{ Inv}_2(S), \ldots\big)
\end{align}
\smallskip\\
Moreover, in all known examples, integral discriminant is a hypergeometric function \cite{GKZ} of those invariants \cite{IntDisc}.

\paragraph{$\bullet$ Independence of action.}Another symmetry of integral discriminants is peculiar independence of action,
expressing the fact that integral discriminant is unchanged if we integrate any function of $S$ instead of exponential:

\begin{align}
Z_{n|r}(S) = \int\limits_{\gamma} d^n x e^{-S(x_1, \ldots, x_n)} = \int\limits_{\gamma^{\prime}} d^n x f\big( \ S(x_1, \ldots, x_n) \ \big), 
\end{align}
\smallskip\\
except for potentially up to a change of integration contour. 

\paragraph{$\bullet$ Vertical symmetry.} Yet another interesting symmetry is vertical symmetry, expressed as

\begin{align}
Z_{n|kr}(S^{k}) = Z_{n|r}(S)
\end{align}
\smallskip\\
Not included in \cite{IntDisc}, however, is an important chapter in the study of integral discriminants -- namely, the study of \emph{correlators} associated with the partition function $Z_{n|r}(S)$:

\begin{align}
\big< \ F(x_1, \ldots, x_n) \ \big> = \dfrac{ \int\limits_{\gamma} d^n x \ F(x_1, \ldots, x_n) \ e^{-S(x_1, \ldots, x_n)} }{ \int\limits_{\gamma} d^n x \ e^{-S(x_1, \ldots, x_n)} }
\end{align}
\smallskip\\
It is the aim of this paper to complete this missing step. We assume that the numerator and denominator have the same integration contour $\gamma$. Correlators can be as simple as a pair correlator of the $x$ variables,

\begin{align}
\big< \ x_i x_j \ \big> = \dfrac{ \int\limits_{\gamma} d^n x \ x_i x_j \ e^{-S(x_1, \ldots, x_n)} }{ \int\limits_{\gamma} d^n x \ e^{-S(x_1, \ldots, x_n)} }
\end{align}
\smallskip\\
which is proportional to $S^{-1}_{ij}$ for the Gaussian case $r=2$. However, given the invariance properties of integral discriminants, of special interest are \emph{invariant correlators}: that is, $\big< \ F(S; x_1, \ldots, x_n) \ \big>$ such that

\begin{align}
F({\widetilde S}; {\widetilde x}_1, \ldots, {\widetilde x}_n) = F(S; x_1, \ldots, x_n), \ \ \ {\widetilde S}_{i_1, \ldots, i_r} = \sum\limits_{j_1, \ldots, j_r = 1}^{n} S_{j_1, \ldots, j_r} U_{i_1, j_1} \ldots U_{i_r, j_r}, \ \ \ {\widetilde x}_{i} = \sum\limits_{j = 1}^{n} U^{-1}_{i, j} x_j
\end{align}
\smallskip\\
for any $U \in SL(n)$. If this condition is met, $SL(n)$ invariance guarantees that $\big< \ F(S; x_1, \ldots, x_n) \ \big>$ is an invariant function of $S$. In this paper, we study this class of invariant correlators. A simple example of an observable $F(S; x_1, \ldots, x_n)$ that meets the above criteria is

\begin{align}
F(S; x_1, \ldots, x_n) = Q(x_1, \ldots, x_n) = \det\limits_{i,j} \dfrac{\partial^2 S(x_1, \ldots, x_n)}{\partial x_i \partial x_j}
\end{align}
\smallskip\\
and we will present more complicated examples below.

Among other phenomena, in this paper we point out the following remarkable property: in several cases invariant correlators turn out to be just polynomial invariants! One of the simplest non-trivial examples of this phenomenon is 

\begin{align}
n|r = 2|4: \ \ \  \big< \ Q(x_1,x_2)^2 \ \big> = \mbox{const} \cdot (2 S_{1111} S_{2222} - 8 S_{1112} S_{1222} + 6 S_{1122}^2)
\end{align}
\smallskip\\
Needless to say, with proper generalization this property could be very useful in the study of non-Gaussian integrals. It is actually reminiscent of \emph{superintegrability} property \cite{Superint} of random matrix models, where in many cases correlators of special polynomials (Schur polynomials, etc.) happen to be again the same polynomials evaluated at special values of parameters. We hope that future studies will shed more light on superintegrability of integral discriminants; the goal of current paper is to work out just the first relevant examples.

\section{The case $n|r = 2|3$}

The simplest possible non-Gaussian case is provided by $n=2$ and $r=3$. Then

\begin{align}
 S(x_1, x_2) = \sum\limits_{i_1, i_2, i_3 = 1}^{2} S_{i_1, i_2, i_3} x_{i_1} x_{i_2} x_{i_3} = S_{111} x_1^3 + 3 S_{112} x_1^2 x_2 + 3 S_{122} x_1 x_2^2 + S_{222} x_2^3
\end{align}
\smallskip\\
In this case, there exists only one algebraically independent polynomial invariant, of degree 4

\begin{align}
& \nonumber I_4(S) = \mathop{\sum\limits_{i_1, i_2, i_3, j_1,j_2,j_3,}^{2}}_{k_1,k_2,k_3,l_1,l_2,l_3 = 1} S_{i_1,i_2,i_3} S_{j_1,j_2,j_3} S_{k_1,k_2,k_3} S_{l_1,l_2,l_3}
\epsilon_{i_1,j_1}
\epsilon_{i_2,j_2}
\epsilon_{k_1,l_1}
\epsilon_{k_2,l_2}
\epsilon_{i_3,k_3}
\epsilon_{j_3,l_3}
\\ & = 2 S_{111}^2 S_{222}^2 - 12 S_{111} S_{112} S_{122} S_{222} + 8 S_{111} S_{122}^3 + 8 S_{112}^3 S_{222} - 6 S_{112}^2 S_{122}^2 = D_{2|3}(S)
\end{align}
\smallskip\\
which simultaneously is the algebraic discriminant of $S$. Note that here we use the completely antisymmetric tensor $\epsilon$, defined by $\epsilon_{11} = \epsilon_{22} = 0, \epsilon_{12} = -\epsilon_{21} = 1$. It was derived in \cite{IntDisc} that

\begin{align}
Z_{2|3}(S) = \int\limits_{\gamma} dx_1 dx_2 e^{-S(x_1, x_2)} = \dfrac{\mbox{const}_{\gamma}}{I_4(S)^{1/6}}
\end{align}
\smallskip\\
We are now interested in computing the correlators of the simplest covariant observable

\begin{align}
& \nonumber Q(x_1, x_2) = \det\limits_{1 \leq i,j \leq 2} \dfrac{\partial^2 S(x_1, x_2)}{\partial x_i \partial x_j} = \sum\limits_{i_1,i_2 = 1}^{2} Q_{i_1,i_2} x_{i_1} x_{i_2} \\
& = 36 (S_{111} S_{122} - S_{112}^2) x_1^2 + 36 (S_{111}S_{222} - S_{112} S_{122}) x_1 x_2 + 36 (S_{112} S_{222} - S_{122}^2) x_2^2
\end{align}
\smallskip\\
namely, the integrals

\begin{align}
\big< \ Q^m \ \big> = \dfrac{1}{Z_{2|3}(S)} \int\limits_{\gamma} dx_1 dx_2 Q(x_1,x_2)^m e^{-S(x_1, x_2)}
\label{Qmcorrelators23}
\end{align}
\smallskip\\
It is easy to show that $\big< \ Q^m \ \big>$ is a homogeneous function of $S$ of degree $4m/3$:

\begin{align}
& \nonumber \big< \ Q^m \ \big>\Big|_{S \rightarrow \lambda S}  = \dfrac{1}{Z_{2|3}( \lambda S)} \int dx_1 dx_2 Q(x_1,x_2)^m e^{-\lambda S(x_1, x_2)} \\
& \nonumber \mathop{=}^{x_i \rightarrow \lambda^{-1/3} y_i} \int \lambda^{-2/3} dy_1 dy_2 \lambda^{-2 m/3 + 2 m} Q(y_1,y_2)^m e^{-S(y_1, y_2)} \\
& = \dfrac{\lambda^{(4m-2)/3}}{\lambda^{-2/3} Z_{2|3}(S)} \int dy_1 dy_2 Q(y_1,y_2)^m e^{-S(y_1, y_2)} = \lambda^{4m/3} \big< \ Q^m \ \big>
\end{align}
\smallskip\\
Since $\big< \ Q^m \ \big>$ is also an invariant function of $S$, we must have

\begin{align}
\big< \ Q^m \ \big> = I_4(S)^{m/3}
\end{align}
\smallskip\\
what concludes our analysis. Remarkably, for all $m$ divisible by 3, $\big< \ Q^m \ \big>$ is a polynomial. This gives a first illustration of the polynomiality property that continues to higher, more complicated, cases.

\section{The case $n|r = 2|4$}

The next-to-simplest non-Gaussian case is provided by $n=2$ and $r=4$. Then

\begin{align}
\hspace{-2ex} S(x_1, x_2) = \sum\limits_{i_1, i_2, i_3, i_4 = 1}^{2} S_{i_1, i_2, i_3, i_4} x_{i_1} x_{i_2} x_{i_3} x_{i_4} = S_{1111} x_1^4 + 4 S_{1112} x_1^3 x_2 + 6 S_{1122} x_1^2 x_2^2 + 4 S_{1222} x_1 x_2^3 + S_{2222} x_2^4
\end{align}
\smallskip\\
In this case, there exist two algebraically independent invariants of $S$, of degrees 2 and 3:

\begin{align}
I_2(S) = 2 S_{1111} S_{2222} - 8 S_{1112} S_{1222} + 6 S_{1122}^2
\end{align}
\begin{align}
I_3(S) = 6 S_{1111} S_{1122} S_{2222} - 6 S_{1111} S_{1222}^2 - 6 S_{1112}^2 S_{2222} + 12 S_{1112} S_{1122} S_{1222} - 6 S_{1122}^3
\end{align}
\smallskip\\
and the algebraic discriminant of $S$ is expressed as a polynomial in these invariants:

\begin{align}
D_{2|4}(S) = I_2^3 - 6 I_3^2
\end{align}
\smallskip\\
It was derived in \cite{IntDisc} that

\begin{align}
Z_{2|4}(S) = \int\limits_{\gamma} dx_1 dx_2 e^{-S(x_1, x_2)} = c^{(1)}_{\gamma} \ I_2^{-1/4} {}_{2}F_{1}\left( \begin{array}{ccc} \frac{1}{12} \ \ \ \frac{5}{12} \\ \frac{1}{2} \end{array} \Big| \dfrac{6 I_3^2}{I_2^3} \right) + c^{(2)}_{\gamma} \ I_3 I_2^{-7/4} {}_{2}F_{1}\left( \begin{array}{ccc} \frac{7}{12} \ \ \ \frac{11}{12} \\ \frac{3}{2} \end{array} \Big| \dfrac{6 I_3^2}{I_2^3} \right)
\label{Branches24}
\end{align}
\smallskip\\
We are now interested in computing the correlators of the simplest covariant observable

\begin{align}
& \nonumber Q(x_1, x_2) = \det\limits_{1 \leq i,j \leq 2} \dfrac{\partial^2 S(x_1, x_2)}{\partial x_i \partial x_j} = \sum\limits_{i_1,i_2,i_3,i_4 = 1}^{2} Q_{i_1,i_2,i_3,i_4} x_{i_1} x_{i_2} x_{i_3} x_{i_4} \\
& \nonumber = 144 (S_{1111} S_{1122} - S_{1112}^2) x_1^4 + 288 (S_{1111} S_{1222} - S_{1112} S_{1122}) x_1^3 x_2 \\
& \nonumber + 144 (S_{1111} S_{2222} + 2 S_{1112} S_{1222} - 3 S_{1122}^2) x_1^2 x_2^2 + 288 (S_{1112} S_{2222} - S_{1122} S_{1222}) x_1 x_2^3 \\
& + 144 (S_{1122} S_{2222} - S_{1222}^2) x_2^4
\end{align}
\smallskip\\
namely, the integrals

\begin{align}
\big< \ Q^m \ \big> = \dfrac{1}{Z_{2|4}(S)} \int\limits_{\gamma} dx_1 dx_2 Q(x_1,x_2)^m e^{-S(x_1, x_2)}
\label{Qmcorrelators24}
\end{align}
\smallskip\\
It is important to note at this point that $Q(x_1, x_2)$ is not the only possible covariant observable. In the case $n|r = 2|4$, it becomes possible to construct a second covariant observable. Namely, instead of taking a determinant of the tensor of second derivatives, we take discriminant of the tensor of third derivatives:

{\fontsize{8pt}{0pt}{\begin{align}
& \nonumber {\widetilde Q}(x_1, x_2) = D_{2|3}\left( {\widetilde S}_{ijk} = \dfrac{\partial^3 S(x_1, x_2)}{\partial x_i \partial x_j \partial x_k} \right) = \sum\limits_{i_1,i_2,i_3,i_4 = 1}^{2} {\widetilde Q}_{i_1,i_2,i_3,i_4} x_{i_1} x_{i_2} x_{i_3} x_{i_4} \\
& \nonumber = -27 (S_{1111}^2 S_{1222}^2 - 6 S_{1111} S_{1112} S_{1122} S_{1222} + 4 S_{1111} S_{1122}^3 + 4 S_{1112}^3 S_{1222} - 3 S_{1112}^2 S_{1122}^2) x_{1}^4 \\
& \nonumber - 54 (S_{1111}^2 S_{1222} S_{2222} - 3 S_{1111} S_{1112} S_{1122} S_{2222} - 2 S_{1111} S_{1112} S_{1222}^2 + 3 S_{1111} S_{1122}^2 S_{1222} + 2 S_{1112}^3 S_{2222} - S_{1112} S_{1122}^3) x_{1}^3 x_{2}  \\
& \nonumber - 27 (S_{1111}^2 S_{2222}^2 - 2 S_{1111} S_{1112} S_{1222} S_{2222} - 6 S_{1111} S_{1122}^2 S_{2222} + 6 S_{1111} S_{1122} S_{1222}^2 + 6 S_{1112}^2 S_{1122} S_{2222} - 8 S_{1112}^2 S_{1222}^2  \\ 
& \nonumber + 6 S_{1112} S_{1122}^2 S_{1222} - 3 S_{1122}^4) x_{1}^2 x_{2}^2 - 54 (S_{1111} S_{1112} S_{2222}^2 - 3 S_{1111} S_{1122} S_{1222} S_{2222} + 2 S_{1111} S_{1222}^3 - 2 S_{1112}^2 S_{1222} S_{2222} \\
& \nonumber + 3 S_{1112} S_{1122}^2 S_{2222} - S_{1122}^3 S_{1222}) x_{1} x_{2}^3 - 27 (S_{1112}^2 S_{2222}^2 - 6 S_{1112} S_{1122} S_{1222} S_{2222} + 4 S_{1112} S_{1222}^3 \\
& + 4 S_{1122}^3 S_{2222} - 3 S_{1122}^2 S_{1222}^2) x_{2}^4
\end{align}}}
\smallskip\\
At first glance, it appears that correlators $\left< Q^{n_1} {\widetilde Q}^{n_2} \right>$ are more general and need to be considered separately. However, it is sufficient to quote a simple relation 

\begin{align}
{\widetilde Q} = \frac{9}{2} I_3 S - \frac{3}{32} I_2 Q
\end{align}
\smallskip\\
so ${\widetilde Q}$ is not really independent. Therefore

\begin{align}
\left< Q^{n_1} {\widetilde Q}^{n_2} \right> = \sum\limits_{m = 0}^{n_2} \dfrac{n_2!}{m!(n_2-m)!} \left( \frac{9}{2} I_3 \right)^m \left( - \frac{3}{32} I_2 \right)^{n_2 - m} \ \left< Q^{n_1 + n_2 - m} S^{m} \right>
\end{align}
\smallskip\\
Correlators with $S$ are easy to compute using the scaling argument:

\begin{align}
& \nonumber \left< Q^{n_1 + n_2 - m} S^{m} \right> = \dfrac{1}{Z_{2|4}(S)} \int dx_1 dx_2 Q(x_1,x_2)^{n_1 + n_2 - m} S(x_1,x_2)^{m} e^{-S(x_1, x_2)} \\
& \nonumber = (-1)^m \dfrac{\partial^m}{\partial \lambda^m} \dfrac{1}{Z_{2|4}(S)} \int dy_1 dy_2 Q(x_1,x_2)^{n_1 + n_2 - m} e^{-\lambda S(x_1, x_2)} \Big|_{\lambda=1} \\
& \nonumber \mathop{=}^{x_i \rightarrow \lambda^{-1/4} y_i} (-1)^m \dfrac{\partial^m}{\partial \lambda^m} \dfrac{1}{Z_{2|4}(S)} \int \lambda^{-1/2} dy_1 dy_2 \lambda^{-(n_1 + n_2 - m)} Q(y_1,y_2)^{n_1 + n_2 - m} e^{-S(y_1, y_2)}\Big|_{\lambda=1} \\
& = \left( (-1)^m \dfrac{\partial^m}{\partial \lambda^m} \lambda^{-(n_1 + n_2 - m + 1/2)} \right) \Big|_{\lambda=1} \left< Q^{n_1 + n_2 - m} \right> = (2)^{-m} \dfrac{(2n_1 + 2n_2 - 1)!!}{(2n_1 + 2n_2 - 2m - 1)!!} \left< Q^{n_1 + n_2 - m} \right>
\end{align}
\smallskip\\
Therefore, we only need to compute correlators (\ref{Qmcorrelators24}), i.e. $\left< Q^{m} \right>$.

To compute these latter correlators, it is not sufficient to use just simple scaling properties, like in the previous $n|r = 2|3$ case. Here, we suggest an important method of computation that relies on the action of the following differential operator:

\begin{align}
{\widehat O}[Q] = \sum\limits_{i_1, i_2, i_3, i_4 = 1}^{2} Q_{i_1,i_2,i_3,i_4} \dfrac{\partial}{\partial S_{i_1,i_2,i_3,i_4}}
\end{align}
\smallskip\\
on the partition function. For the actual computation, let us fix a basis for invariant series

\begin{align}
I_2^{d} \omega^{n}, \ \ \ \ \ \omega = \dfrac{I_3^2}{I_2^3}
\end{align}
\smallskip\\
In this basis, we find the following formulas for the action of this operator:

\begin{align}
{\widehat O}[Q] \ I_2^{d} \omega^{n} = (144d - 432n) I_2^{d+1/2} \omega^{n+1/2} + 72 n I_2^{d+1/2} \omega^{n-1/2}
\end{align}
or in the invariant form,

\begin{align}
{\widehat O}[Q] = 144 I_2^{3/2} \omega^{1/2} \dfrac{\partial}{\partial I_2} - 432 I_2^{1/2} \omega^{3/2} \dfrac{\partial}{\partial \omega} + 72 I_2^{1/2} \omega^{1/2} \dfrac{\partial}{\partial \omega}
\end{align}
\smallskip\\
It is easy to derive that

\begin{align}
{\widehat O}[Q] \cdot S = Q, \ \ \ {\widehat O}[Q] \cdot Q = 1728 I_2 S
\end{align}
\smallskip\\
Denoting for simplicity $Z_{2|4}(S) = Z$, we find

\begin{align}
\nonumber {\widehat O}[Q] \cdot Z \big< Q^{n_1} \big> = {\widehat O}[Q] \cdot \int Q^{n_1} e^{-S} = \int e^{-S} \big( - Q^{n_1} {\widehat O}[Q] \cdot S + n_1 Q^{n_1-1} {\widehat O}[Q] \cdot Q \big)  \\
= - Z\big< Q^{n_1+1} \big> + 1728 n_1 I_2 Z\big< Q^{n_1-1} S \big> = - Z\big< Q^{n_1+1} \big> + 864 n_1 (2 n_1 - 1) I_2 Z\big< Q^{n_1-1} \big>
\end{align}
\smallskip\\
Using the following ansatz for the correlator

\begin{align}
Z \left< Q^{n_1} \right> = I_2^{\frac{2 n_1 - 1}{4}} \omega^{\frac{n_1 {\rm mod} 2}{2}} \sum\limits_{n = 0}^{\infty} C_{n_1}(n) \omega^n
\end{align}
\smallskip\\
we find that coefficients satisfy the recursion

\begin{align}
\boxed{ \ \ \ \begin{array}{lll}
\nonumber C_{n_1+1}(n) = - 72 \left(n+1-\frac{n_1 {\rm mod} 2}{2}\right) C_{n_1}(n + 1 - n_1 {\rm mod} 2) \\
\emph{} - ( - 432 n + 216 n_1 {\rm mod} 2 + 72 n_1 - 36 \big) C_{n_1}(n - n_1 {\rm mod} 2) + 864 n_1 (2 n_1-1) C_{n_1-1}(n)
\end{array}
\ \ \ }
\end{align}
\smallskip\\
For the first branch of the integral discriminant (\ref{Branches24}), i.e. the one associated with $c^{(2)}_{\gamma} = 1, c^{(2)}_{\gamma} = 0$, the initial condition can be seen as 

\begin{align}
C_0(n) = \mbox{const} \cdot 6^n \dfrac{\Gamma(n + 1/12)\Gamma(n + 5/12)}{\Gamma(n + 1/2) n!}
\end{align}
\smallskip\\
Also, it is assumed that all $C_{n_1}(n) = 0$ for $n_1 < 0$. With this information, we can give the general solution:

\begin{align}
\dfrac{C_{1}(n)}{C_0(n)} = \dfrac{6(1+12 n)}{1 + 2n}, \ \ \ \dfrac{C_{2m + 1}(n)}{C_0(n)} = \dfrac{-48 \ 2592^m \ \Gamma\left(2 m + \frac{3}{2}\right) \Gamma\left(\frac{1}{4} + m - n\right) \Gamma\left(\frac{3}{4} - 3 n\right) }{ (2 n + 1) \Gamma\left(\frac{1}{2}\right) \Gamma\left(\frac{1}{4} - n\right) \Gamma\left(m - 3 n - \frac{1}{4}\right)}, \ m \geq 1
\label{Correlator24case1}
\end{align}
\begin{align}
\dfrac{C_{2m}(n)}{C_0(n)} = \dfrac{2592^m \ \Gamma\left(2 m + \frac{1}{2}\right) \Gamma\left(\frac{1}{4} + m - n\right) \Gamma\left(\frac{3}{4} - 3 n\right) }{ \Gamma\left(\frac{1}{2}\right) \Gamma\left(\frac{1}{4} - n\right) \Gamma\left(m - 3 n + \frac{3}{4}\right)}, \ m \geq 0
\label{Correlator24case2}
\end{align}
\smallskip\\
This summarizes our solution for correlators associated with the first branch of the integral discriminant. For second branch, consideration is similar. Again, we would like to draw attention to the following fact: in some cases, the correlator happens to give a polynomial invariant. This is most simply seen in the case $\left< Q^{2} \right>$ which corresponds to $m = 1$ in (\ref{Correlator24case2}): then, $\dfrac{C_{2}(n)}{C_0(n)} = 648$ is independent of $n$. Therefore,

\begin{align}
\left< Q^{2} \right> = \dfrac{ I_2^{\frac{3}{4}} }{Z} \sum\limits_{n = 0}^{\infty} C_{2}(n) \omega^n = \dfrac{ 648 I_2^{\frac{3}{4}} }{Z} \sum\limits_{n = 0}^{\infty} C_{0}(n) \omega^n = 648 I_2
\end{align}
\smallskip\\
and we discover once again that non-Gaussian correlator equals a polynomial invariant.

Another interesting observation about eqs. (\ref{Correlator24case1}) and (\ref{Correlator24case2}) is that particular correlator made of $Q$ vanishes:

\begin{align}
\left< I_3 Q^{2m} + \dfrac{1}{31104 (4m+1)(2m+1)(4m+3)} Q^{2m+3} - \dfrac{(4m+3)I_2}{48(2m+1)(4m+1)} Q^{2m+1} \right> = 0
\end{align}
\smallskip\\

\section{The case $n|r = 3|3$}

Yet another tractable non-Gaussian case is provided by $n=3$ and $r=3$. Then

\begin{align}
& \nonumber S(x_1, x_2, x_3) = \sum\limits_{i_1, i_2, i_3 = 1}^{3} S_{i_1, i_2, i_3} x_{i_1} x_{i_2} x_{i_3} = S_{111} x_1^3 + 3 S_{112} x_1^2 x_2 + 3 S_{113} x_1^2 x_3 + 3 S_{122} x_1 x_2^2 \\
& + 6 S_{123} x_1 x_2 x_3 + 3 S_{133} x_1 x_3^2 + S_{222} x_2^3 + 3 S_{223} x_2^2 x_3 + 3 S_{233} x_2 x_3^2 + S_{333} x_3^3
\end{align}
\smallskip\\
In this case, there exist two algebraically independent invariants of $S$, of degrees 4 and 6:

\begin{align}
I_4(S) = \mathop{\sum\limits_{i_{11},i_{12},i_{13},i_{21},i_{22},i_{23}}^{3}}_{i_{31},i_{32},i_{33},i_{41},i_{42},i_{43} = 1}  S_{i_{11},i_{21},i_{31}} S_{i_{12},i_{22},i_{43}} S_{i_{13},i_{42},i_{32}} S_{i_{41},i_{23},i_{33}} \epsilon_{i_{11},i_{12},i_{13}} \epsilon_{i_{21},i_{22},i_{23}} \epsilon_{i_{31},i_{32},i_{33}} \epsilon_{i_{41},i_{42},i_{43}}
\end{align}
\begin{align}
I_6(S) = \mathop{\sum\limits_{i_{11},i_{12},i_{13},i_{21},i_{22},i_{23},i_{31},i_{32},i_{33}}^{3}}_{i_{41},i_{42},i_{43},i_{51},i_{52},i_{53},i_{61},i_{62},i_{63} = 1}  \begin{array}{ll} S_{i_{11},i_{21},i_{62}} S_{i_{31},i_{22},i_{53}} S_{i_{12},i_{32},i_{43}} S_{i_{13},i_{41},i_{61}} S_{i_{33},i_{42},i_{51}} S_{i_{63},i_{23},i_{52}} \\ \epsilon_{i_{11},i_{12},i_{13}} \epsilon_{i_{21},i_{22},i_{23}} \epsilon_{i_{31},i_{32},i_{33}} \epsilon_{i_{41},i_{42},i_{43}} \epsilon_{i_{51},i_{52},i_{53}} \epsilon_{i_{61},i_{62},i_{63}} \end{array}
\end{align}
\smallskip\\
Note that here we use the completely antisymmetric tensor $\epsilon$, defined by $\epsilon_{123} = \epsilon_{231} = \epsilon_{312} = -\epsilon_{213} = -\epsilon_{321} = -\epsilon_{132} = 1$ and $\epsilon_{ijk} = 0$ otherwise. The algebraic discriminant of $S$ is simply expressed through these invariants\footnote{The normalization of $I_4$ is different from the one used in \cite{IntDisc}, explaining the discrepancy in coefficient with that paper.}:

\begin{align}
D_{3|3}(S) = I_4^3 + 6 I_6^2
\end{align}
\smallskip\\
In terms of these variables, 

\begin{align}
\hspace{-6ex} Z_{3|3}(S) = \int\limits_{\gamma} dx_1 dx_2 d x_3 e^{-S(x_1, x_2, x_3)} = c^{(1)}_{\gamma} \ I_6^{-1/6} {}_{2}F_{1}\left( \begin{array}{ccc} \frac{1}{12} \ \ \ \frac{7}{12} \\ \frac{2}{3} \end{array} \Big| \dfrac{- I_4^3}{6 I_6^2} \right) + c^{(2)}_{\gamma} \ I_4 I_6^{1/2} {}_{2}F_{1}\left( \begin{array}{ccc} \frac{5}{12} \ \ \ \frac{11}{12} \\ \frac{4}{3} \end{array} \Big| \dfrac{- I_4^3}{6 I_6^2} \right)
\label{Branches33}
\end{align}
\smallskip\\
We are now interested in computing the correlators of the simplest covariant observable

{\fontsize{8pt}{0pt}{
\begin{align}
& \nonumber Q(x_1, x_2, x_3) = \det\limits_{1 \leq i,j \leq 3} \dfrac{\partial^2 S(x_1, x_2)}{\partial x_i \partial x_j} = \sum\limits_{i_1,i_2,i_3,i_4 = 1}^{2} Q_{i_1,i_2,i_3} x_{i_1} x_{i_2} x_{i_3} \\
& \nonumber = 216 (S_{111} S_{122} S_{133} - S_{111} S_{123}^2 - S_{112}^2 S_{133} + 2 S_{112} S_{113} S_{123} - S_{113}^2 S_{122}) x_1^3 \\
& \nonumber + 216 (S_{111} S_{122} S_{233} - 2 S_{111} S_{123} S_{223} + S_{111} S_{133} S_{222} - S_{112}^2 S_{233} + 2 S_{112} S_{113} S_{223} - S_{112} S_{122} S_{133} + S_{112} S_{123}^2 - S_{113}^2 S_{222}) x_1^2 x_2 \\
& \nonumber 
+ 216 (S_{111} S_{122} S_{333} - 2 S_{111} S_{123} S_{233} + S_{111} S_{133} S_{223} - S_{112}^2 S_{333} + 2 S_{112} S_{113} S_{233} - S_{113}^2 S_{223} - S_{113} S_{122} S_{133} + S_{113} S_{123}^2) x_1^2 x_3 \\
& \nonumber 
+ 216 (S_{111} S_{222} S_{233} - S_{111} S_{223}^2 - S_{112} S_{122} S_{233} + S_{112} S_{133} S_{222} + 2 S_{113} S_{122} S_{223} - 2 S_{113} S_{123} S_{222} - S_{122}^2 S_{133} + S_{122} S_{123}^2) x_1 x_2^2 \\
& \nonumber 
+ 216 (S_{111} S_{222} S_{333} - S_{111} S_{223} S_{233} - S_{112} S_{122} S_{333} - 2 S_{112} S_{123} S_{233} + 3 S_{112} S_{133} S_{223} + 3 S_{113} S_{122} S_{233} - 2 S_{113} S_{123} S_{223} - S_{113} S_{133} S_{222} \\
& \nonumber 
 - 2 S_{122} S_{123} S_{133} + 2 S_{123}^3) x_1 x_2 x_3 + 216 (S_{112} S_{222} S_{233} - S_{112} S_{223}^2 - S_{122}^2 S_{233} + 2 S_{122} S_{123} S_{223} - S_{123}^2 S_{222}) x_2^3 \\
& \nonumber 
+ 216 (S_{111} S_{223} S_{333} - S_{111} S_{233}^2 - 2 S_{112} S_{123} S_{333} + 2 S_{112} S_{133} S_{233} + S_{113} S_{122} S_{333} - S_{113} S_{133} S_{223} - S_{122} S_{133}^2 + S_{123}^2 S_{133}) x_1 x_3^2 \\
& \nonumber 
+ 216 (S_{112} S_{222} S_{333} - S_{112} S_{223} S_{233} + S_{113} S_{222} S_{233} - S_{113} S_{223}^2 - S_{122}^2 S_{333} + 2 S_{122} S_{133} S_{223} + S_{123}^2 S_{223} - 2 S_{123} S_{133} S_{222}) x_2^2 x_3 \\
& \nonumber 
+ 216 (S_{112} S_{223} S_{333} - S_{112} S_{233}^2 + S_{113} S_{222} S_{333} - S_{113} S_{223} S_{233} - 2 S_{122} S_{123} S_{333} + 2 S_{122} S_{133} S_{233} + S_{123}^2 S_{233} - S_{133}^2 S_{222}) x_2 x_3^2 \\
& \nonumber 
+ 216 (S_{113} S_{223} S_{333} - S_{113} S_{233}^2 - S_{123}^2 S_{333} + 2 S_{123} S_{133} S_{233} - S_{133}^2 S_{223}) x_3^3
\end{align}}}
\smallskip\\
namely, the integrals

\begin{align}
\big< \ Q^m \ \big> = \dfrac{1}{Z_{3|3}(S)} \int\limits_{\gamma} dx_1 dx_2 dx_3 Q(x_1,x_2,x_3)^m e^{-S(x_1, x_2, x_3)}
\label{Qmcorrelators33}
\end{align}
\smallskip\\
Note that, unlike the case $n|r = 2|4$, the choice of discriminant of the tensor of third derivatives is not easily available here, because it would be independent of $x$ and thus not so interesting in correlators. For the actual computation, we use the same method based on the action of the differential operator

\begin{align}
{\widehat O}[Q] = \sum\limits_{i_1, i_2, i_3 = 1}^{3} Q_{i_1,i_2,i_3} \dfrac{\partial}{\partial S_{i_1,i_2,i_3}}
\end{align}
\smallskip\\
on the partition function. For the actual computation, let us fix a basis for invariant series

\begin{align}
I_6^{d} \rho^{n}, \ \ \ \ \ \rho = \dfrac{I_4^3}{I_6^2}
\end{align}
\smallskip\\
In this basis, we find the following formulas for the action of this operator:

\begin{align}
{\widehat O}[Q] \ I_6^{d} \rho^{n} = (-36 d + 72 n) I_6^{d+1/3} \rho^{n+2/3} + 432 n I_6^{d+1/3} \rho^{n-1/3}
\end{align}
or in the invariant form,

\begin{align}
{\widehat O}[Q] = - 36 I_6^{4/3} \rho^{2/3} \dfrac{\partial}{\partial I_6} + 72 I_6^{1/3} \rho^{5/3} \dfrac{\partial}{\partial \rho} + 432 I_6^{1/3} \rho^{2/3} \dfrac{\partial}{\partial \rho}
\end{align}
\smallskip\\
It is easy to derive that

\begin{align}
{\widehat O}[Q] \cdot S = Q, \ \ \ {\widehat O}[Q] \cdot Q = -648 I_4 S
\end{align}
\smallskip\\
Denoting for simplicity $Z_{3|3}(S) = Z$, we find

\begin{align}
\nonumber {\widehat O}[Q] \cdot Z \big< Q^{n_1} \big> = {\widehat O}[Q] \cdot \int Q^{n_1} e^{-S} = \int e^{-S} \big( - Q^{n_1} {\widehat O}[Q] \cdot S + n_1 Q^{n_1-1} {\widehat O}[Q] \cdot Q \big)  \\
= - Z\big< Q^{n_1+1} \big> - 648 n_1 I_4 Z\big< Q^{n_1-1} S \big> = - Z\big< Q^{n_1+1} \big> - 648 n_1^2 I_4 Z\big< Q^{n_1-1} \big>
\end{align}
\smallskip\\
where we used the fact that $\big< Q^{n_1-1} S \big> = n_1 \big< Q^{n_1-1} \big>$ for $n|r = 3|3$. Using the following ansatz

\begin{align}
Z \left< Q^{n_1} \right> = I_6^{\frac{2 n_1 - 1}{6}} \rho^{\frac{(-n_1) {\rm mod} 3}{3}} \sum\limits_{n = 0}^{\infty} C_{n_1}(n) \rho^n
\end{align}
\smallskip\\
we find that coefficients satisfy the recursion

{\fontsize{8pt}{0pt}{
\begin{align}
\boxed{ \ \ \ 
\left\{ \begin{array}{lll}
C_{n_1+1}(n) = (12 n_1 - 72 n + 42) C_{n_1}(n-1) - (432 n + 144) C_{n_1}(n) - 648 n_1^2 C_{n_1-1}(n-1), \ \ \ (-n_1) \mod 3 = 1 \\
C_{n_1+1}(n) = (12 n_1 - 72 n + 18) C_{n_1}(n-1) - (432 n + 288) C_{n_1}(n) - 648 n_1^2 C_{n_1-1}(n), \ \ \ (-n_1) \mod 3 = 2 \\
C_{n_1+1}(n) = (12 n_1 - 72 n - 6) C_{n_1}(n) - (432 n + 432) C_{n_1}(n+1) - 648 n_1^2 C_{n_1-1}(n), \ \ \ (-n_1) \mod 3 = 0
\end{array} \right.
\ \ \ }
\end{align}}}
\smallskip\\
For the first branch of the integral discriminant (\ref{Branches33}), i.e. the one associated with $c^{(2)}_{\gamma} = 1, c^{(2)}_{\gamma} = 0$, the initial condition can be seen as 

\begin{align}
C_0(n) = \mbox{const} \cdot \left(\dfrac{-1}{6}\right)^n \dfrac{\Gamma(n + 1/12)\Gamma(n + 7/12)}{\Gamma(n + 2/3) n!}
\end{align}
\smallskip\\
Also, it is assumed that all $C_{n_1}(n) = 0$ for $n_1 < 0$. With this information, the general solution for $C_{n_1}(n)$ is determined uniquely. Unlike the case $n|r = 2|4$, in this case we were not able to give a closed form solution in terms of Gamma functions. To give one example,

\begin{align}
\dfrac{C_6(n)}{C_0(n)} = \dfrac{77396705280 (81 n^2 + 495 n + 176)}{(12 n - 5)(12 n - 11)}
\end{align}
\smallskip\\
contains a non-factorizable polynomial in the numerator.

A few comments about this solution are in order. First, it is interesting how the study of correlators distinguishes the two cases $n|r = 2|4$ and $3|3$, which at the level of partition functions are in complete parallel. Namely, the recursion for correlators depends on divisibility by 2 for $n|r = 2|4$ and on divisibility by 3 for $n|r = 3|3$. Second, the phenomenon of polynomiality continues to present itself: here,

\begin{align}
\dfrac{C_2(n)}{C_0(n)} = -432
\end{align}
\smallskip\\
and is independent of $n$, therefore by the same argument as in the previous section, 

\begin{align}
\left< Q^{2} \right> = \dfrac{ I_6^{\frac{1}{2}} \rho^{1/3} }{Z} \sum\limits_{n = 0}^{\infty} C_{2}(n) \omega^n = \dfrac{ -432 I_6^{\frac{-1}{6}} I_4 }{Z} \sum\limits_{n = 0}^{\infty} C_{0}(n) \omega^n = -432 I_4
\end{align}
\smallskip\\
which is a manifestation of polynomiality. This completes our description of the $n|r=3|3$ case.

\section{Conclusion}

In this paper, we continued the systematic description of simplest non-Gaussian models -- integral discriminants -- to the level of correlators. We only consider invariant correlators, of which the simplest representative example is $\left< Q^{m} \right>$ for $Q = \det_{n \times n} \partial^2 S$. A general method is suggested to compute those correlators using the action of certain differential operators on the partition function. Remarkably, in many cases the correlators happen to be very simple functions of $S$ -- polynomial. We think that this \emph{polynomiality property} deserves attention, because it could provide useful for generalization to $n \rightarrow \infty$, i.e. to functional (or path) integrals. 

To give a more detailed comment on that issue, theory of integral discriminants is appealing, beautiful, and in many aspects follows closely the theory of ordinary discriminants. However, the complexity of the answers grows very fast as $n|r$ increase. The results of \cite{IntDisc} seem to suggest that general integral discriminants are A-hypergeometric functions of Gelfand, Kapranov and Zelevinsky \cite{GKZ} of the set of polynomial invariants. This is a rather complicated answer. For this reason, we think this theory would benefit significantly from a small 'wonder', or in other words some simplification. It does not necessarily mean we want to consider a simpler object -- no, we would like to study the same integral discriminants as before. Rather, simplification could be found in a new way of looking at the same theory, or some new angle. The study of correlators may provide just that. 

The 'wonder' here is arguably the polynomiality property, which reduces complexity from hypergeometric functions of invariants to simply invariants themselves. Of course, it remains to be seen if this property continues beyond the simplest cases $n|r = 2|3, 2|4$ and $3|3$ that we considered here. In this direction the case $n|r = 2|5$ is the next plausible candidate for investigation, because of clear hypergeometric structures seen there \cite{IntDisc}. Another possibility is to consider the sequence of cases $n|r = n|3$ where $n$ can be arbitrarily large, and see if polynomiality property can be seen there. This is left for future work.

\section*{Acknowledgements}

The work is supported by the state assignment of the Institute for Information Transmission Problems of RAS.

\end{document}